\documentclass[12pt]{aipproc}
\layoutstyle {6x9}
\usepackage{epsfig}
\SetInternalRegister\hbadness{8000}
\newcommand{\be}{\begin{equation}}
\newcommand{\ee}{\end{equation}}

\newcommand\doingARLO[2][]{%
  \ifx\mmref\undefined #1\else #2\fi
  }
  
\begin{document}
\title{Light-light and heavy-light mesons in the model of QCD
string with quarks at the ends}

\author{A.V.Nefediev}{
  address={ITEP, 117218, B.Cheremushkinskaya 25, Moscow, Russia},
  email={nefediev@heron.itep.ru}
}
\begin{abstract}
The variational einbein field method is applied to the
model of the QCD string with quarks at the ends for the
case of light--light and heavy--light mesons. Special
attention is payed to the proper string dynamics.
The correct string slope of the Regge trajectories is
reproduced for light--light states which comes out from the
picture of rotating string.
Masses of several low-lying orbitally and radially excited
states in the $D$, $D_s$, $B$, and $B_s$ meson spectra are calculated
and a good agreement with the experimental data as well as
with recent lattice calculations is found. The role of the
string correction to the interquark interaction is discussed
at the example of the identification of ${D^*}'(2637)$ state
recently claimed by DELPHI Collaboration.
For the heavy--light mesons the standard constants used in Heavy
Quark Effective Theory are extracted and compared to the results
of other approaches.
\end{abstract}
\date{}
\maketitle

\section{Introduction}

One of the most beautiful phenomena observed in QCD --- namely, the formation
of an
extended string between the colour sources, implies that the string degrees of
freedom in hadrons should be taken into account in the proper way. In the present
contribution the model of the QCD string with quarks at the ends is used to
investigate the spectra of the light--light and heavy--light mesons. The variational
einbein field method (see \cite{ll} and references therein) is used in 
numerical calculations of the spectra.

\section{Light-light mesons}

Starting from the gauge invariant Green's function of a $q\bar q$ meson, 
performing integration in the path integral over the fermionic and gluonic
fields, and using the minimal area law assumption in the latter case, one can 
extract
the Lagrangian of the spinless quark-antiquark system in the form (see $e.g.$
\cite{DKS})
\be
L(t)=-m_1\sqrt{\dot{x}_1^2}-m_2\sqrt{\dot{x}_2^2}
-\sigma\int_0^1d\beta\sqrt{(\dot{w}w')^2-\dot{w}^2w'^2},
\label{1}
\ee
where we synchronize the quark times, $x_{10}=x_{20}=t$, and
choose the minimal-string profile function in the 
straight-line form, $w_{\mu}(t,\beta)=\beta x_{1\mu}(t)+(1-\beta)x_{2\mu}(t)$.

If the einbeins $\mu_{1,2}$ and $\nu(\beta)$ are now introduced in (\ref{1})
to get rid of the square roots, then the centre-of-mass Hamiltonian for the case
of massless quarks reads $(\mu_1=\mu_2=\mu)$
\be
H=\frac{p^2_r}{\mu}+\mu+U(\mu,r),\quad\quad
U(\mu,r)=\frac{\sigma r}{y} \arcsin~y+\mu y^2,
\label{3}
\ee
where the extremum in $\nu$ is already taken at the level of the Hamiltonian
yielding $\nu_{\rm ext}(\beta)=\sigma r(1-4y^2(\beta-\frac{1}{2})^2)^{-1/2}$, 
so that $y$ is the solution to the transcendental equation
\be
\frac{L}{\sigma r^2}=\frac{1}{4y^2}
(\arcsin~y-y\sqrt{1-y^2})+\frac{\mu{y}}{\sigma r}.
\label{4}
\ee

The spectrum of the Hamiltonian (\ref{3}) is found using the quasiclassical
method with the consequent minimization of each eigenvalue with
respect to the einbein $\mu$, which is thus treated as a variational parameter,
playing the role of the effective constituent mass of the quark. It appears
due to the interaction and takes the value of $200-300MeV$ even if one starts
with zero current quark mass.
The results of numerical calculations with $\sigma=0.17GeV^2$ are given in 
Fig.1. The interested
reader can find details in papers \cite{ll}. Note, that the Regge
trajectories for the light--light mesons remain nearly straight-line up to very
low momenta and the only fitting parameter is the overall negative shift
$\Delta M^2\sim -1GeV$, one and the same for all three trajectories.
Another important comment is that the correct string slope of the trajectories,
$2\pi\sigma$, appears quite naturally in the given approach as a consequence of
the rotating string inertia properly taken into account in the Hamiltonian
(\ref{3}), so that the form of the effective potential
$U(\mu,r)$ does not amount to the naive sum of the centrifugal barrier for the
quarks and the linearly rising potential $\sigma r$ \cite{ll}. 

\begin{figure}[t]
\centerline{\epsfig{file=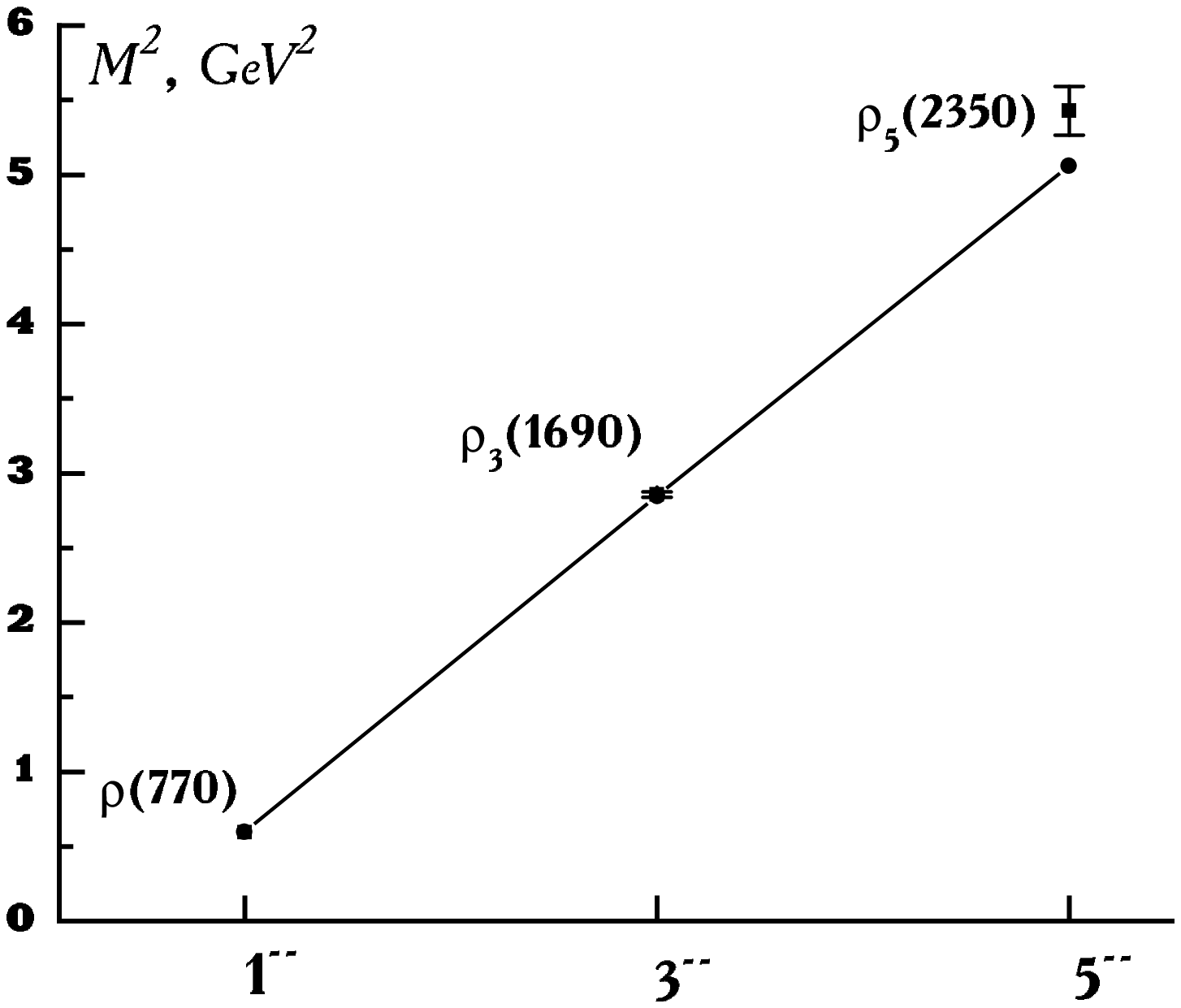,width=6cm}\hspace{-1.5cm}
\epsfig{file=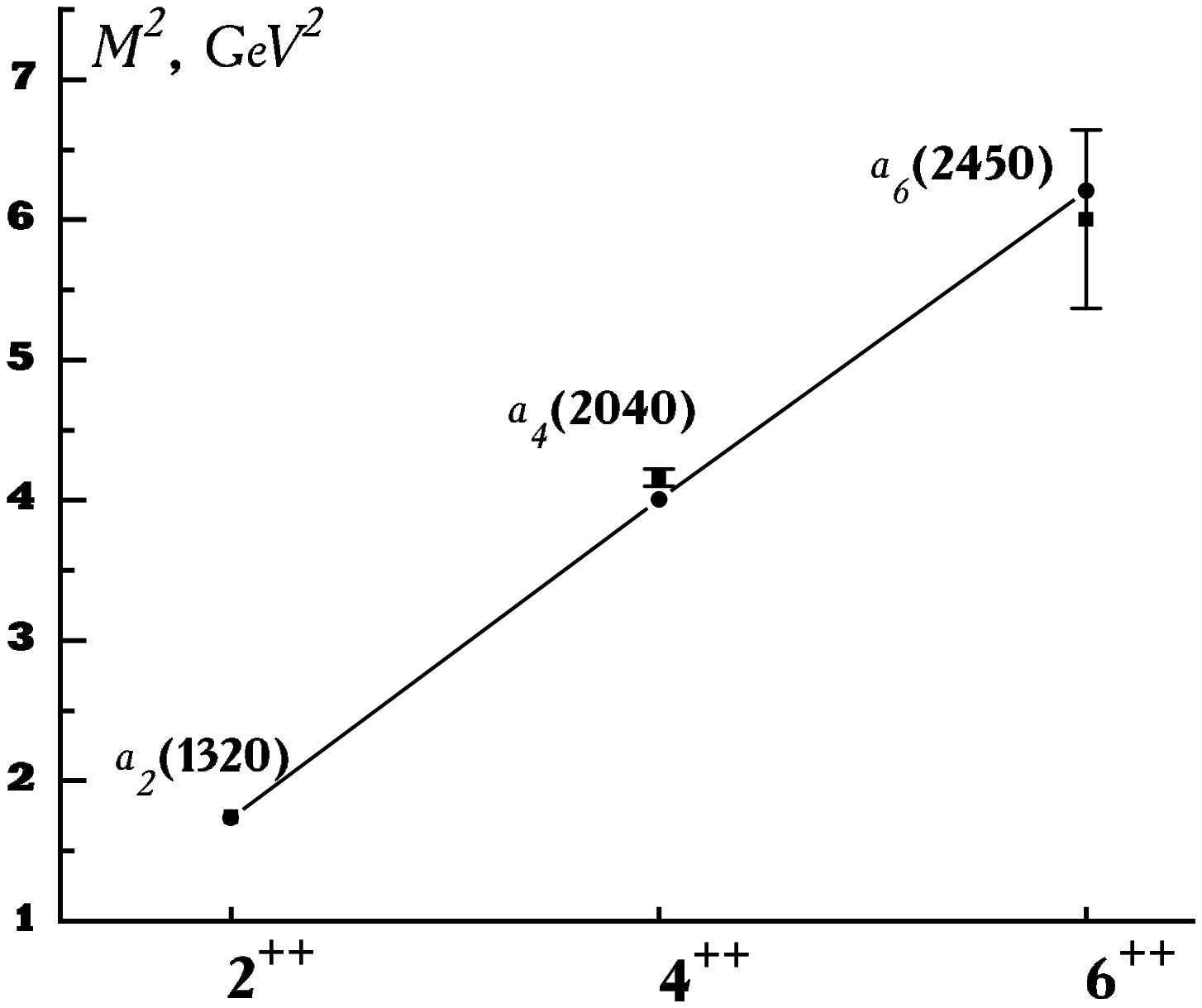,width=6cm}\hspace*{-1.5cm}
\epsfig{file=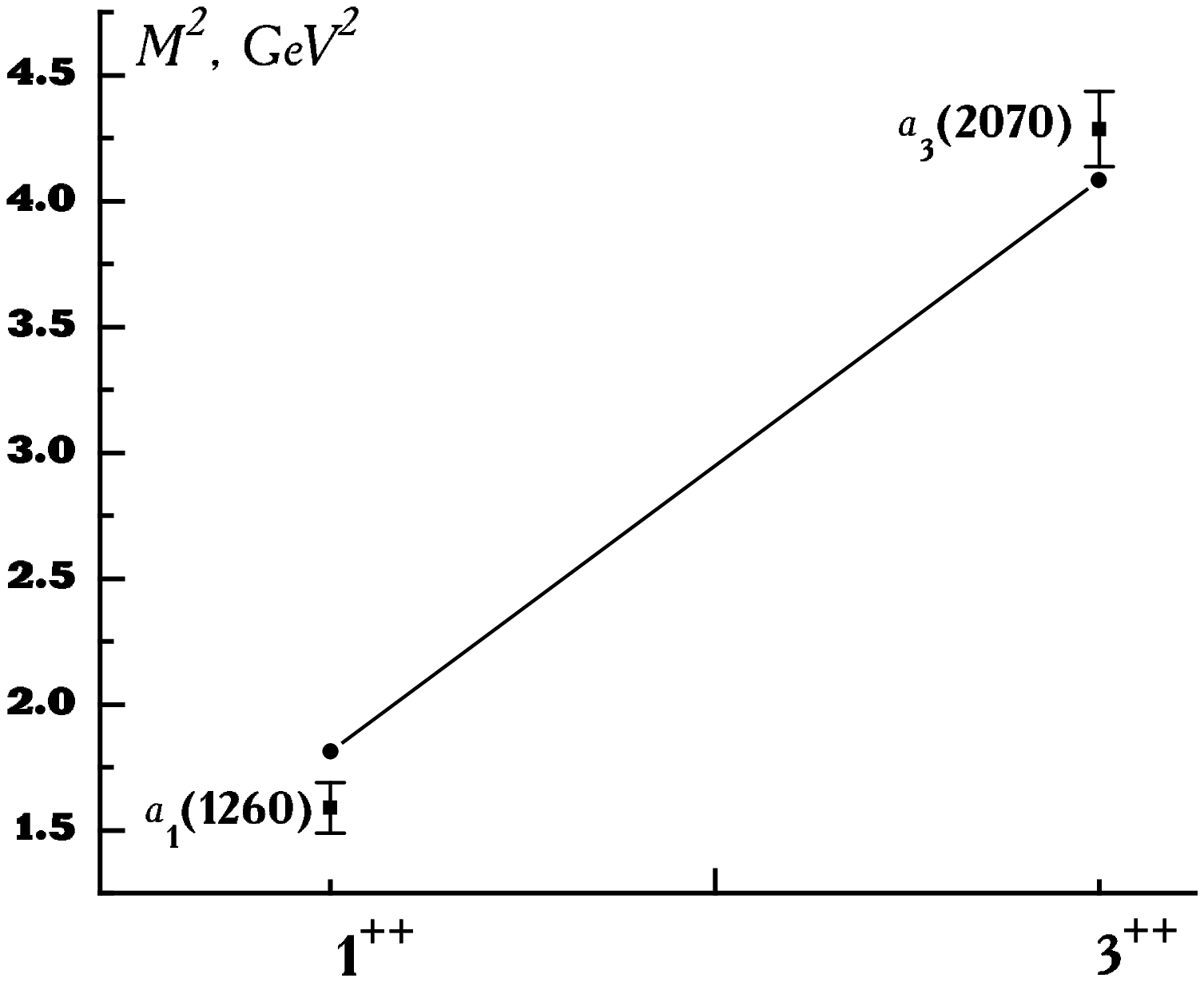,width=6cm}}
\caption{The lowest Regge trajectories for the light--light mesons (experimental 
data are given by boxes with error bars).}
\end{figure}
	    
\section{Heavy-light mesons}
\begin{table}[t]
\begin{tabular}{ccccccccc}
\hline
\bf Splitting&\bf $D_s$--$D$&\bf $D_s^*$--$D^*$&\bf $D^*$--$D$&\bf $D_s^*$--$D_s$&\bf $B_s$--$B$&\bf $B_s^*$--$B^*$&\bf $B^*$--$B$&\bf $B_s^*$--$B_s$\\
\hline
Experiment&99&102&141&144&90&91&46&47\\
\hline
Theory&114&115&146&147&100&102&63&65\\
\hline
\end{tabular}
\caption{\bf Splittings for the $D$, $D_s$, $B$, and $B_s$ mesons in $MeV$.}
\end{table}

In the calculations of the light-light meson spectra we totally ignored the quark
spin, concentrating on the proper account of the string dynamics. Now, 
to have reliable predictions for the heavy--light meson masses, we supply the
Hamiltonian for spinless quarks connected by the string,
\be
H_0=\sum_{i=1}^2\left(\frac{\vec{p}^2+m_i^2}{2\mu_i}+\frac{\mu_i}{2}\right)+\sigma
r-\frac{\sigma
(\mu_1^2+\mu_2^2-\mu_1\mu_2)}{6\mu_1^2\mu_2^2}\frac{\vec{L}^2}{r},
\label{5}
\ee
by spin-dependent corrections due to confining and the OGE interaction
$(\kappa=-\frac43\alpha_s)$,
\be
V_{sd}=\frac{8\pi\kappa}{3\mu_1\mu_2}(\vec{S}_1\vec{S}_2)
\left|\psi(0)\right|^2
+\frac{\kappa}{\mu_1\mu_2r^3}\left(3(\vec{S}_1\vec{n})
(\vec{S}_2\vec{n})-(\vec{S}_1\vec{S}_2)\right)
-\frac{\sigma}{2r}\left(\frac{\vec{S}_1\vec{L}}{\mu_1^2}+
\frac{\vec{S}_2\vec{L}}{\mu_2^2}\right)
\label{6}
\ee
$$
+\frac{\kappa}{r^3}\left(\frac{1}{2\mu_1}+\frac{1}{\mu_2}\right)
\frac{\vec{S}_1\vec{L}}{\mu_1}
+\frac{\kappa}{r^3}\left(\frac{1}{2\mu_2}+\frac{1}{\mu_1}\right)
\frac{\vec{S}_2\vec{L}}{\mu_2}
+\frac{\kappa^2}{2\pi\mu^2r^3}
\left(\vec{S}\vec{L}\right)(1.43-{\rm ln}(\mu r)),
$$
as well as by the Coulomb interaction $-\frac43\frac{\alpha_s}{r}$ and the 
overall negative constant shift $-C_0$. The latter remains the only fitting
parameter, whereas we use the standard values for others: $\sigma=0.17GeV^2$,
$m_u=5MeV$, $m_d=9MeV$, $\alpha_s=0.4$ for $D$ mesons and $\alpha_s=0.39$ for
$B$'s. The last term on the r.h.s. of equation (\ref{5})
is the string correction which is always negative and 
accounts for the proper string dynamics. The
results of numerical calculations for the spectra of orbitally and radially
excited $D$, $D_s$, $B$, and $B_s$ mesons, as well as comparison with the lattice
data and results of other approaches, can be found in \cite{ll,hl}. In 
Table~1 we give the splittings for the above-mentioned mesons compared to the
experimental values. Let us also quote the masses of the radially $(n=1)$ and
orbitally $(l=2)$ excited $D$ mesons: $M(0^-)=2664MeV$, $M(2^-)=2663MeV$, 
$M(3^-)=2654MeV$. Thus the $3^-$ state is the lightest one and it is the most 
probably
candidate for the resonance ${D^*}'(2637)$ recently claimed by the DELPHI
Collaboration \cite{DELPHI}. Such an identification resolves the problem
usually encountered in the framework of the quark models: being too narrow this
meson can not be associated with the first radial excitation, whereas model
predictions for orbitally
excited states lie about $50-60MeV$ higher than needed. In our approach
the proper string dynamics, in the form of the correction to the Hamiltonian, 
lowers the energy of the orbitally excited state $3^-$. It is also instructive to
note, that the fitted value of the parameter $C_0$ is insensitive to the heavy
quark and depends only on the light-quark content of the meson, that supports
the idea that it is due to the self-energy of the latter.

\section{Bridge to Heavy Quark Effective Theory}

The suggested approach allows us to find analytic formulae for the constants
used in the Heavy Quark Effective
Theory, as well as to evaluate them numerically. In the standard
parameterization the mass of a heavy-light meson is
\be
M_{hl}=m_Q+\bar\Lambda-(\lambda_1+d_H\lambda_2)/2m_Q+O\left(1/m_Q^2\right)
\ee
with $d_H$ being +3 for $0^-$ states or -1 for $1^-$ ones. For the idealized
case ($m_1\to\infty,\;m_2=0$) our formulae simplify considerably, giving
analytical expressions for all three constants \cite{ll}. 
A more reliable way to estimate two of them is to find the best fit of the form
\be
M_{fit}=m_Q+\bar\Lambda+C_0-\lambda_1/2m_Q
\label{fit}
\ee
with $C_0$ fixed by fitting the experimental spectrum, as discussed above,
and varying $m_Q$ around the bottom quark mass. The results are listed in 
Table~2, where they are compared with those of other approaches,
demonstrating good agreement with the latter.

\begin{table}[t]
\begin{tabular}{cccccc}
\hline
&\bf $m_1\to\infty$ $m_2\to 0$&\bf Fit (\ref{fit})&\bf Sum rules \cite{sumrules}&\bf $B$ mesons decays \cite{Bdec}&
\bf DS equation \cite{simtjon}\\
\hline
$\bar\Lambda$, $GeV$&0.471&0.485&0.4 $\div$ 0.5&0.39 $\pm$ 0.11&0.493/0.288\\
\hline
$\lambda_1$, $GeV^2$&-0.506&-0.379&-0.52 $\pm$ 0.12&-0.19 $\pm$ 0.10&-\\
\hline
$\lambda_2$, $GeV^2$&0.21&0.17&0.12&0.12&-\\
\hline
\end{tabular}
\caption{\bf Standard parameters used in HQET.}
\end{table}

\section{Conclusions}

In conclusion let us emphasize that the proper dynamics of the gluonic degrees
of freedom in hadrons, taken into account in the form of an effective QCD string
between quarks, are of paramount importance in establishing the hadron spectra
of mass and their properties. Account for the inertia of the rotating 
string inside mesons allowed us to reproduce the correct string slope of the
Regge trajectories, as well as to fit the spectrum of $D$ and $B$ mesons,
and to resolve the problem of identification of the state recently claimed by
the DELPHI Collaboration. The variational einbein field method used in
the calculations is proved to be efficient and accurate, that allows one to use
it in investigations of various relativistic systems.

Financial support of RFFI grants 00-02-17836, 00-15-96786, and 01-02-06273,
INTAS-RFFI grant IR-97-232 and INTAS CALL 2000-110 
is gratefully acknowledged. 

\doingARLO[\bibliographystyle{aipproc}]
          {\ifthenelse{\equal{\AIPcitestyleselect}{num}}
             {\bibliographystyle{arlonum}}
             {\bibliographystyle{arlobib}}
          }
\bibliography{sample}

\end{document}